\documentclass{epsconf}
\usepackage{graphicx}
\usepackage{epsfig} % use this package to include EPS format figures
\usepackage{wrapfig}
\usepackage{amsmath}

\title{Seeding of proton bunch self-modulation by an electron bunch in plasma}
\author{\underline{L. Verra}$^{1,2,3,}\footnote{livio.verra@cern.ch}$    , G. Zevi Della Porta$^1$, K.-J. Moon$^4$, \\A.-M. Bachmann$^2$, E. Gschwendtner$^1$ and P. Muggli$^2$}
\institute{$^1$ CERN, Geneva 1211, Switzerland\\
$^2$ Max Planck Institute for Physics, Munich 80805, Germany\\
$^3$ Technical University Munich, Garching 85748, Germany \\
$^4$ UNIST, Ulsan 44919, South Korea}
\begin{document}
\maketitle
\section{Introduction}
The AWAKE experiment at CERN $\cite{PATRIC:READINESS}$ relies on the self-modulation in plasma of the long $400\,$GeV/c proton bunch from the CERN SPS, to accelerate an externally injected electron bunch to GeV energies. The control of the acceleration requires that the self-modulation process and the electron beam injection are reproducible from event to event. Making the self-modulation instability (SMI) reproducible means that the phase and the amplitude of the plasma wakefields along the driver bunch are fixed, once the process has saturated. This is achieved by seeding the instability, and the process is therefore called seeded self-modulation (SSM). Proton bunch SSM using a relativistic ionization front method $\cite{KARL:PRL,MARLENE:PRL,FABIAN:PRL}$, and the acceleration of electrons $\cite{NATURE}$ were demonstrated during AWAKE Run 1.
%During AWAKE Run 1 (2016-2018), both the seeded self-modulation of the $400\,$GeV proton bunch from CERN SPS $\cite{KARL:PRL, MARLENE:PRL,FABIAN:PRL}$ and the acceleration of electrons up to $2\,$GeV were experimentally demonstrated $\cite{NATURE}$. The self-modulation was seeded using a relativistic ionization front, therefore the head of the proton bunch remained un-modulated. AWAKE Run 2 $\cite{PATRIC:PLAN, EDDA:IPAC}$ will be split in different phases, and in its final versions (Run 2c, 2d) the experiment will consist of two plasma sections: the first one dedicated to the self-modulation of the proton bunch; the second one to the acceleration of the electron bunch. As the plasma in the second section will be pre-formed, it is necessary that the whole long proton bunch self-modulates with reproducible phase in the first section. Otherwise, the unmodulated part of the bunch could undergo self-modulation instability and disrupt the structure of the wakefields. The solution to this issue is to seed the self-modulation using a short electron bunch $\cite{PATRIC:SEEDING}$. 
The physics of seeding using a short electron bunch will be studied during Run 2a (starting in July 2021) $\cite{PATRIC:PLAN,EDDA:IPAC}$.
\section{Electron bunch seeding of the self-modulation}
\begin{wrapfigure}{r}{50mm}
\vspace{0cm} % Adjust vertical figure placement
\includegraphics[width=50mm]{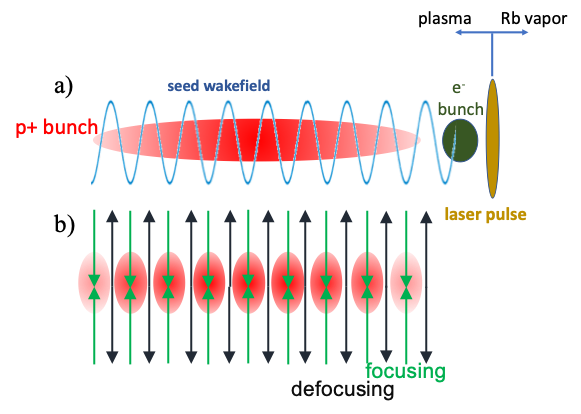}
\caption{\it \small Schematic of the electron bunch seeding process. a) Beams at the injection and initial wakefields; b) Fully modulated train of microbunches and transverse wakefields at saturation.}
\label{fig:essm}
\vspace{0cm} % Adjust vertical figure spacing
\end{wrapfigure}

%When electrons travel in plasma, they drive wakefields that modulates the plasma electron density. If this modulations is deep enough, it will impose a modulation of the proton bunch charge density. Again, if this is deep enough, the self-modulation process is triggered.
When an electron bunch travels in plasma, it drives wakefields that can impose a charge modulation on the trailing proton bunch (see Figure $\ref{fig:essm}$a). If this modulation is deep enough (i.e. the amplitude of the wakefields is above the seeding threshold), the self-modulation process is seeded and grows resonantly until the proton bunch is fully modulated (see Figure $\ref{fig:essm}$b). Thus, the final phase of the microbunch train (and of the wakefields) is uniquely related in phase to the short seed electron bunch.
\par AWAKE Run 2a is using the same experimental setup as Run 1 $\cite{PATRIC:READINESS, EDDA:IPAC}$. %The same electron source and transfer line $\cite{KEVIN:SOURCE, CHIARA:LINE}$ that were used to provide the electron beam for external injection and acceleration is now used to generate the seed electron beam. 
The main elements are: a $10\,$meter-long Rb vapor source, a $120\,$fs, $<450\,$mJ laser pulse
($\lambda = 780\,$nm) ionizing the Rb atoms and creating the plasma column, the electron beam source and transfer line, a magnetic spectrometer system downstream of the plasma. The initial plasma electron density $n_{pe}$ can be varied in a range from 0.5 to $10\cdot 10^{14}\,$cm$\textsuperscript{-3}$, the energy of the electron bunch from $10$ to $20\,$MeV with $0.5\%\,$relative energy spread, the charge from $100$ to $600\,$pC (with normalized transverse emittance and bunch length $\sigma_z$ scaling accordingly from 1 to $6\,$mm$\cdot$mrad and from 2 to $5\,$ps). The electron bunch transverse size at the plasma entrance can be adjusted using a quadrupole triplet final focusing system to a minimum size $\sigma_r\sim200\,\mu$m.
\par
%%
%\begin{figure}[!h]\centering
%\end{figure}
%%
As shown in $\cite{FABIAN:PRL}$, the phase of the train of proton microbunches is reproducible (therefore SSM has occurred) when the amplitude of the seed wakefield is larger than a threshold value. Using the relativistic ionization front seeding method, the threshold was determined to be between 4 and $6\,$MV/m (with $n_{pe} = 0.9\cdot 10^{-14}\,$cm$\textsuperscript{-3}$).
Linear plasma wakefields theory $\cite{LIN_THEORY}$ shows that with the initial parameters of the AWAKE electron bunch, it would not be possible to effectively seed the self-modulation. For $Q = 150\,$pC, $\sigma_z=2\,$ps, ${\sigma_r=200 \, \mu}$m, ${n_{pe} = 2\cdot 10^{-14}\,}$cm${\textsuperscript{-3}}$, the maximum amplitude of the transverse wakefields $W_{\perp}$ behind the bunch at $r=\sigma_r$ is $3\,$MV/m. The electron beam transverse size, though, evolves according to $\gamma m \frac{d^2 \sigma_r}{dt^2} = q W_{\perp}$ (where $q$ and $m$ are the electron charge and mass and $\gamma$ the relativistic factor): 
using linear optics, one  can estimate the initial bunch radius $\sigma_{r0}$ at the injection so that the amplitude of the wakefields within the bunch at $r=\sigma_{r0}$ balances the divergence of the bunch and therefore the size remains constant along the plasma. For the same parameters mentioned above, $\sigma_{r0}\sim 35 \,\mu$m, much smaller than the minimum achievable size at the plasma entrance in the experiment. Therefore, the bunch undergoes severe non-linear pinching in the first centimeters of propagation in plasma $\cite{KOSTANTIN:EVOLUTION}$, and numerical simulations are required to describe and compute its transverse size.
Simulations $\cite{ANNA:THESIS, KJ:SIM}$ show that, because of this transverse size evolution, the bunch charge density becomes high enough to drive transverse wakefields with amplitude above the seeding threshold. 
%Moreover, the transverse size remains small until the amplitude of the wakefields decreases due to longitudinal dephasing, making the beam length comparable to the plasma wavelength and causing charge loss.

%(Figures from KJ showing pinching, amplitude of the wakefields, radial charge distribution at point of minimum size). 
The goal of Run 2a is not only to experimentally prove the seeding of the self-modulation of the long proton bunch using an electron bunch, but also to vary the initial parameters of the electron bunch (and therefore the amplitude of the initial wakefields) to observe the transition between SMI and SSM.
\section{Preparatory experimental studies}
\begin{figure}[!h]\centering
%\vspace{0cm} % Adjust vertical figure placement
\includegraphics[width=160mm]{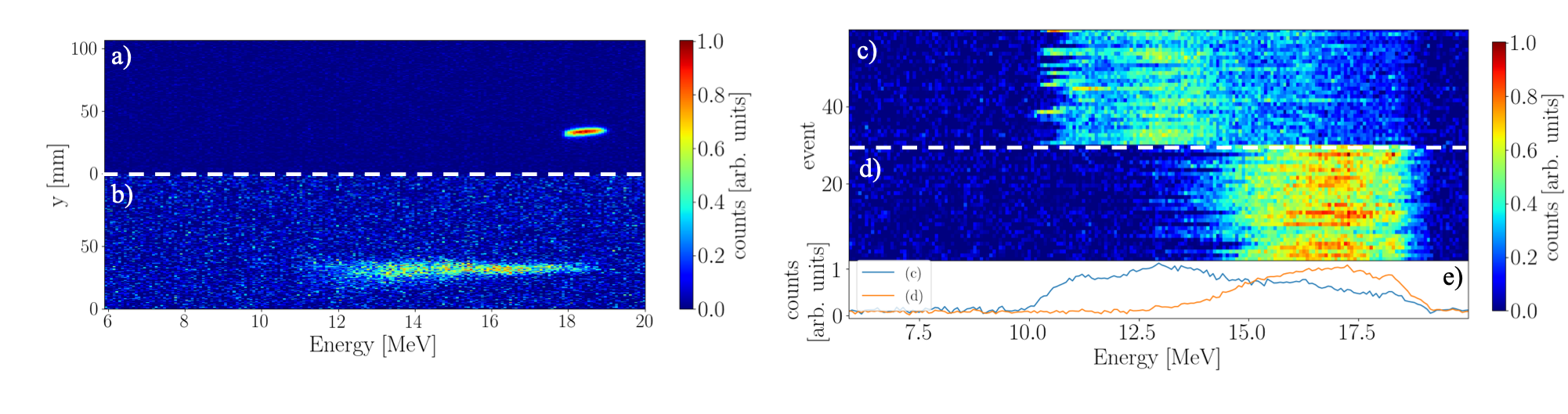}\centering
\caption{\it \small a) Electron bunch imaged onto the spectrometer screen after propagation through vacuum. b) Electron bunch imaged onto the spectrometer screen after propagation through $10\,$ meters of plasma. The horizontal axis is the dispersive plane. The counts of each image are normalized to the respective maximum. c) and d) show a waterfall plot of the energy projections of the electron beam imaged on the spectrometer screen. c) 30 events with the smallest beam size at the plasma entrance; d) 30 events with the largest beam size at the plasma entrance. e) sums of the energy distributions for the two cases: (c) in blue and (d) in orange. For all images, the electron bunch charge at the plasma entrance is $150\,$pC and the energy is $\sim18.5\,$MeV.}
\label{fig:waterfall}
%\vspace{0cm} % Adjust vertical figure spacing
\end{figure}

Before performing the electron bunch seeding experiment with the SPS proton bunch, we want to study the effect of the propagation through the 10 meter-long plasma column $\cite{GIO:IPAC}$ on the electron bunch. As the low-energy bunch drives wakefields and loose a significant amount of its energy, we expect a fraction of the electrons to be dephased with respect to the wakefields, and to be defocused, and to detect the energy loss as a long low-energy tail in the energy spectrum of the bunch on the spectrometer screen.
Figure $\ref{fig:waterfall}$a shows the $150\,$pC and $\sim18.5\,$MeV electron bunch, propagated through vacuum and imaged onto the spectrometer screen. Figure $\ref{fig:waterfall}$b shows the electron bunch with the same input parameters after propagation through $10\,$m of plasma ($n_{pe} = 2\cdot 10^{14}\,$cm$\textsuperscript{-3}$). As the beam is dispersed in the horizontal plane, the horizontal axis of the screen is converted into energy. One notices the long low-energy tail, showing energy loss occurred in the plasma.
\par During the first experimental campaign, we varied the size at the plasma entrance of the $150\,$pC, $18.5\,$MeV electron bunch (and $n_{pe} = 2\cdot 10^{14}\,$cm$\textsuperscript{-3}$). Figure $\ref{fig:waterfall}$ (c and d) show a waterfall plot of the projections along the energy axis of images at the spectrometer screen. We collected 30 events with the smallest beam size at the injection (beam focused at the plasma entrance, $\sigma_r \sim 200\,\mu$m (c)) and 30 events with the largest beam size (beam focused $5\,$m downstream of the plasma entrance, $\sigma_r \sim 1\,$mm (d)). We note here that the energy distributions reach lower values for the smaller beam size (the minimum energy is detected around $10\,$MeV) indicating that the smaller beam (with larger charge density) experiences more energy loss.
We also show in Figure $\ref{fig:waterfall}$e the sum of the energy distributions for the two cases: the minimum energy and the mean of the distribution are clearly different in the two cases. 
The energy of one electron along the propagation length $L$ evolves according to: $\Delta W = q\int_{0}^{L} E_z(z) dz$, where $E_z$ is the longitudinal wakefield within the bunch. What we observe on the screen is the results of a complex dynamics, as $E_z$ varies along the bunch and along the plasma.  
\par
%%
%\begin{figure}[!h]\centering
%\begin{wrapfigure}{r}{50mm}\centering
%\vspace{0cm} % Adjust vertical figure placement
%\includegraphics[width=50mm]{AWAKEdryrun_sim_wakefield.png}
%\caption{\it \small Will be: Ez and Wr vs. z to show that they are high in first 2 meters}
%\label{fig:wfields}
%\vspace{0cm} % Adjust vertical figure spacing
%\end{wrapfigure}
%\end{figure}
%%
Simulations $\cite{KJ:SIM}$ show
%(Figure $\ref{fig:wfields}$) 
that the amplitude of the decelerating fields inside the bunch and of the transverse fields behind the bunch are maximum over the first $2\,$meters of plasma, where the transverse pinching occurs. Then, due to dephasing, some charge is expelled out of the plasma and the bunch length becomes comparable to the plasma wavelength, making the wakefields decrease and the bunch transverse size increase.
%Afterwards, the electrons are defocused (and even if they are detected on the spectrometer screen, they do not lose more energy) or they are slightly accelerated.
%Therefore, a difference in energy loss is attributable in the amplitude of the wakefields in that region. 
%For the remaining length of the plasma, the least energetic particles start dephasing respect to the wakefields. Thus, they end up in the defocusing phase (and they are expelled out of the plasma) or in the accelerating phase, where they slowly recover some energy.
To summarize, the energy spectrum, observed after the propagation in plasma, is mostly affected by the longitudinal and transverse evolution of the bunch and of the wakefields over the first meters of propagation, which in turn are directly related to the initial parameters of the electron bunch, in particular its charge density.
%This could be a signature for larger amplitude wakefields. We would therefore have the ability of varying the amplitude of the seed wakefields by changing the initial parameters of the electron bunch, allowing us to observe the transition between SMI and SSM of the proton bunch during Run 2a.
%This will be very important during Run 2a to observe the transition between SMI and SSM of the proton bunch.
\section{Conclusions}
We have discussed the electron bunch seeding process of the proton bunch self-modulation, that will be studied in the context of AWAKE Run 2a. 
%We were able to detect variation of the energy loss (and therefore of the wakefields amplitude) when changing the initial parameters of the seed electron bunch. 
We have shown that we are able to vary the energy loss of the seed electron bunch in plasma by changing its initial parameters. We will perform the same type of measurements in the presence of the proton bunch, including the variation of other parameters (e.g. electron bunch charge, for a fixed bunch size), to study the phase reproducibility of the self-modulation of the proton bunch and the transition from SMI to SSM, using the electron bunch seeding method.

\end{document}